\documentclass[lettersize,journal]{IEEEtran}
\usepackage{amsmath,amsfonts}
\usepackage{algorithmic}
\usepackage{algorithm}
\usepackage{array}
\usepackage[caption=false,font=normalsize,labelfont=sf,textfont=sf]{subfig}
\usepackage{textcomp}
\usepackage{stfloats}
\usepackage{url}
\usepackage{verbatim}
\usepackage{graphicx}
\usepackage{cite}
\usepackage{moreverb,url}

\usepackage[colorlinks,bookmarksopen,bookmarksnumbered,citecolor=red,urlcolor=red]{hyperref}
\usepackage{amsmath,amssymb,amsfonts,mathtools}
\usepackage{algorithmic}
\usepackage{graphicx}
\usepackage{textcomp}
\usepackage{xcolor}
\usepackage[capitalise,nameinlink,noabbrev]{cleveref} 
\crefname{equation}{}{}
\creflabelformat{equation}{#2(#1)#3}
\usepackage{tikz}
\usepackage{array}
\usetikzlibrary{arrows, decorations.markings}
\usetikzlibrary{decorations.pathreplacing}
\usetikzlibrary{decorations.text}
\usetikzlibrary{decorations.pathmorphing}
\usetikzlibrary{arrows}
\usetikzlibrary{patterns}
\usetikzlibrary{arrows.meta}

\usepackage{lineno,hyperref}
\modulolinenumbers[5]

\begin{document}

\title{Design and Simulation-Based Testing of Connected Traffic Light Guidance Systems}

\author{Michael Khayyat$^{*}$\thanks{*Corresponding author.}, Alberto Gabriele, Francesca Mancini, Stefano Arrigoni, and Francesco Braghin,~\IEEEmembership{Member,~IEEE}
\thanks{All authors are with the Department of Mechanical Engineering, Politecnico di Milano (Via La Masa, 1, 20156 Milano MI, Italy) \{michael.khayyat, alberto.gabriele, francesca.mancini, stefano.arrigoni, francesco.braghin\}@polimi.it }}%

\markboth{IEEE TRANSACTIONS ON INTELLIGENT VEHICLES,~Vol.~xx, No.~x, Month~202x}%
{Shell \MakeLowercase{\textit{et al.}}: A Sample Article Using IEEEtran.cls for IEEE Journals}


\maketitle

\begin{abstract}
The establishment of fast and reliable communication technologies, such as $5$G, is enabling the evolution of a new generation of connected ADAS. This work aims to develop a traffic light advisory system, Multiple Traffic Light Advisor (MTLA), to improve driving efficiency and intersection viability, and reduce urban pollution. The developed system guides the driver on how to modify the vehicle speed to efficiently utilize the current and future states of the traffic lights ahead. Starting from a non-optimal implementation of the overall architecture, MTLA is further improved through a non-linear MPC approach. The developed system is tested in a virtual environment in IPG CarMaker and results show good performances with a high potential and space for future developments.
\end{abstract}


\section{Introduction}

Automotive research has identified environmental pollution as a critical issue. Road transportation is a major contributor to air pollution, and addressing this issue involves addressing challenges such as travel delays and fuel consumption. To promote sustainable transportation, vehicle manufacturers have implemented various solutions, including the use of lighter and stronger materials, alternative fuels, and more efficient powertrain components. Advanced Driver Assistance Systems (ADAS) also offer promising solutions for reducing fuel consumption and emissions \cite{EMISSION1,EMISSION2}.\newline
One system designed to improve both safety and sustainability in transportation is called GLOSA (Green Light Optimal Speed Advisory). Its goal is to reduce emissions and fuel consumption by minimizing the number of stops at intersections. To achieve this, GLOSA calculates a recommended speed profile and provides warnings or control actions to adjust the vehicle's velocity. GLOSA systems can be divided into two categories based on the number of traffic lights they consider in real-time to provide a recommended speed: Single-segment (S-GLOSA) or Multiple segment (M-GLOSA). S-GLOSA systems only analyze the first traffic light ahead of the vehicle, while M-GLOSA systems take into account multiple traffic lights along the vehicle's route. There are various approaches in literature for calculating the recommended velocity profile for S-GLOSA systems. In \cite{STL1}, Barth et al. interpret the eco-driving strategy as a minimization of the total tractive power demand and idling time. For instance, to avoid idling, the vehicle should get to the traffic light during its green phase. Thus, considering $t_r$ and $t_g$ as the time until phase changes respectively to red and green, an admissible time interval to get the green is defined as in \cref{STL1_t}: 
\begin{equation}
  t\in  
\begin{cases}
[0, t_r)\cup[t_g,t_{r,1}) & \text{if phase = green}\\
[t_g,t_r) & \text{if phase = red}
\end{cases}
\label{STL1_t}
\end{equation}
Where $t_r$ is the time to the first green to red shift while $t_{r,1}$ is the time to the second green to red shift. Corresponding admissible velocities are defined considering a constant velocity profile ($[V_{lo},V_{ho}]$). Consistency with the road limits is checked as in \cref{STL1_v} and the minimum and maximum possible velocities (respectively $V_l$ and $V_h$) are defined. Then, the maximum one is suggested to the driver.
\begin{equation}
    V_\text{possible} = [V_{lo},V_{ho}]\cap[0,V_\text{limit}]=[V_l,V_h]
    \label{STL1_v}
\end{equation}
In \cite{STL2}, Cai and Ning propose a speed guidance system considering two different algorithms: ``Best Feasibility” and ``Best Efficiency”. The first focuses on reducing driver annoyance from the suggested speed by having the smallest difference between actual and guidance speed. ``Best Efficiency" algorithm aims at passing the traffic light as fast as possible. It is concluded that the optimal strategy is given by a combination of both algorithms: ``Best Feasibility" during braking phases while ``Best Efficiency" in high speed and acceleration situations.\newline
Katsaros et al. \cite{STL5} study the impacts of GLOSA on fuel and traffic efficiency by analyzing average fuel consumption and average stop time behind a traffic light. The target velocity is computed as following: the time needed for the vehicle to reach the traffic light is computed considering a uniformly accelerated motion profile, then if the vehicle reaches the traffic light when it is green, maximum road speed is suggested to the driver, otherwise the target speed is computed considering a uniformly accelerated motion so to reach the traffic light during the next green phase. Simulations show that, in a high traffic density scenario, the higher the number of equipped vehicles, the higher the benefits. On the other hand, traffic efficiency increases if traffic density decreases.\newline
When dealing with M-GLOSA systems, two main design approaches can be distinguished: Model Predictive Control (MPC) and Genetic Algorithms (GAs). For the application under analysis, traffic light phase changes have to be considered as constraints. Unfortunately, phases dynamic variability makes the feasible solution space non-convex, resulting in a computationally expensive optimization problems which may not converge to global optimum \cite{MPC_MTL1}. This issue is solved by managing the problem on two levels: a lower level takes into account for phase variations and defines a first target velocity, then the latter is given as reference to a higher MPC based level, that computes the final desired velocity. The resulting solution may be sub-optimal, but is real-time implementable.\newline
Asadi and Vahidi \cite{MPC_MTL1} propose a ``Predictive Cruise Control (PCC)” that minimizes the use of brakes based on traffic signal information and enforcing at the same time several physical constraints. A set of logical rules calculates a reference velocity for timely arrival at green lights considering a constant velocity profile. The obtained profile is fed as reference to the MPC that tracks this target velocity. The controller uses a vehicle model that is based on the linearization of the longitudinal dynamics and takes into account for vehicle mass and position, aerodynamic drag, rolling resistance and road grade forces. The authors define the cost function as to minimize brake force and deviation from target speed. Constraints bound speed, engine/brake forces and safe distance between follower and lead vehicle. Tests results show $59\%$ reduction of fuel consumption, $39\%$ less $CO_2$ emission and reduced travel time when the PCC controller is adopted. Further simulations are performed in \cite{MPC_MTL2} to prove the effective reduction of fuel consumption and travel time of the developed system.\newline
In \cite{MPC_MTL4}, Jones et al. utilize an MPC approach to control an  electric vehicle approaching a road segment with multiple traffic lights. Energy-optimal and time saving trajectories are computed considering a lower level ``Fast MPC" to compute a first attempt trajectory and a ``Main MPC” which computes a more detailed trajectory, taking into account for the vehicle dynamic behavior and acceleration/deceleration limits. A linear kinematic vehicle model is used. The cost function minimizes the desired acceleration (input of the system) and deviation of vehicle trajectory with respect to the reference one. It is noteworthy that minimizing the desired acceleration is an equivalent way of minimizing energy consumption of the vehicle, since the torque required to the electric motor is directly connected to acceleration.Constraints are adopted to limit vehicle position, speed and desired acceleration.\newline
In \cite{GA1, GA3}, an advisory speed is proposed to the driver according to selected preferences like minimisation of total traveling time or fuel consumption. Testing shows that in free-flow conditions such multi-segment GLOSA gives much better results when compared with single-segment approach. Nguyen et al. \cite{GA2} propose an improved GLOSA method called R-GLOSA, which also takes into account traffic density to compute the optimal speed. Density information is obtained through the vehicle communication with Road Side Units (RSU) distributed along the road. Both single and multiple R-GLOSA are developed and compared with single/multiple GLOSA and no-GLOSA vehicle; results shows that the developed approach is better than non-RSU ones in terms of travel time and waiting time and that $CO_2$ emissions are reduced according to vehicle density.\newline
It is important to highlight that even though there are many GLOSA system in literature, none of them, to the best of our knowledge, address the following issues: comfort, variability of friction coefficient and minimization of setup variation with respect to a standard vehicle in order to apply the system. In this paper, a novel ADAS which tackles the issues above, is developed and discussed.\newline
The developed guidance system, named Traffic Light Advisor (TLA), warns the driver in time on how to modify the vehicle velocity to get one (Single Traffic Light Advisor) or more (Multiple Traffic Light Advisor) green traffic lights. To this end, TLA utilizes connectivity for obtaining necessary information, such as traffic light phases, road geometry and friction coefficient and speed limits. The means of connectivity in this work is 5G, due to its high speed and reliability. Two different versions of this ADAS are presented, a non-optimal MTLA and an optimal MTLA. The first one aims to keep a hardware configuration as coherent as possible with the one available in standard commercial vehicles. The second adopts Model Predictive Control techniques to improve comfort, which is crucial for such applications, at the expense of computational power required.
\section{Setup Description}
\subsection{System Architecture}
\begin{figure}[t]
	\centering
	\includegraphics[keepaspectratio=true,width = 0.75\linewidth]{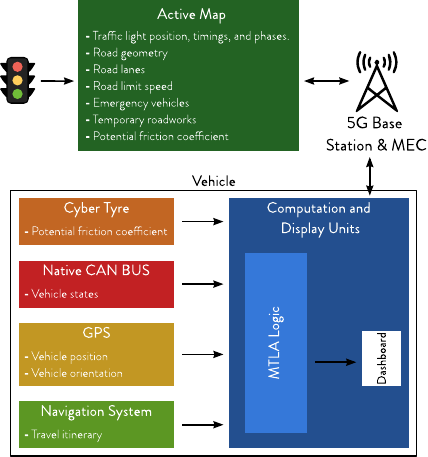}
	\caption{Intelligent Speed Adaptation \& Control - Data Transmission Logic Scheme}
	\label{UC21}
\end{figure}
Traditional V2V technology consists of wireless data transmissions between vehicles. Commonly, I2V communications are wireless, bidirectional, and similarly to V2V, use Dedicated Short-Range Communication (DSRC) frequencies to transfer data \cite{VehicleCommunication}.\newline
In this work, a new approach is used. Information is not transmitted directly between vehicles or between vehicle and infrastructure, but an intermediate and dynamic layer, that is in charge of the management of it. Thanks to 5G and edge computing features of low latency, high band, the possibility to connect millions of devices and to elaborate data where they are received, a real application of it is possible. In particular, Multiaccess Edge Computing (MEC) manages all communication flows and Active Map population. 
\cref{UC21} demonstrates how the map is populated with information from the traffic light (timing and phases) and friction information. Road potential grip information is obtained through smart tyres; once measured, this is transmitted to the Active Map, along with the corresponding GPS location: in this way the map is populated with this data. Traffic light timing and phases, and the potential friction are part of the dynamic information stored in the map. Together with dynamic information, static ones such as road geometry, road lanes, road limit speed and traffic lights position is present in the map too. This is labelled as static since it does \textit{not change} in time. The TLA algorithms need further input quantities (other than the information retrieved from the Active Map). As seen in \cref{UC21}, additional inputs are given by the vehicle CAN BUS, GPS and navigation system. Furthermore, the vehicle is equipped with a Computation and Display Unit, a tablet device which has the function to perform calculations required by the TLA logic and display any warnings or advice to the driver.\newline
In \cref{INP_TL}, the input quantities for both TLA systems are reported.
\begin{table}[]
	\centering
	\caption{Algorithm input parameter classification}
	\label{INP_TL}
	\scriptsize
	\begin{tabular}{|c|c|c|c|}
		\hline
		\textbf{Parameter} &  \textbf{Variability} & \textbf{Source} & \textbf{Type}\\
		\hline
		$V$& Dynamic& CAN BUS& Input\\
		\hline
		$X$ &  Dynamic& GPS& Input\\
		\hline
		$Y$&  Dynamic& GPS& Input\\
		\hline
	    $\Psi$&  Dynamic& GPS& Input\\
		\hline
		$\mu$&  Dynamic& Active Map& Input/Output\\
		\hline
		Road Geometry&  Static& Active Map & Input\\
		\hline
		Road Limit Speed&  Static/Dynamic& Active Map & Input\\
		\hline
		Traffic Light Position &  Static/Dynamic& Active Map & Input\\
		\hline
		Traffic Light Phases &  Dynamic& Active Map & Input\\
		\hline
		Traffic Light Timing &  Dynamic& Active Map & Input\\
		\hline
	\end{tabular}
\end{table}
Speed, position and orientation ( defined according to GPS reference system) of the vehicle, friction coefficient, limit speed and geometry of the road are obtained as previously described. It is noteworthy to mention that for these applications the road geometry is considered as the path to follow and is used only to relate the states ($X,Y,\Psi$) to the abscissa $s$ of the path, while information about curves is disregarded.%
\subsection{General Algorithm - Traffic Light Advisor Overview}
\label{TLA_overview}

\begin{figure}
	\centering 
    \includegraphics[keepaspectratio=true,width=\linewidth]{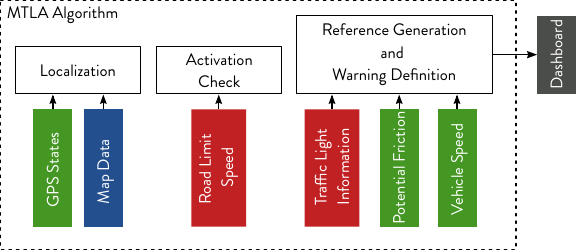}
	\caption{TLA algorithm scheme. Blocks colored in green, blue, and red correspond to dynamic, hybrid, or static data respectively.}
	\label{TLA_scheme}
\end{figure}

The MTLA considers with the following scenarios:
\begin{itemize}
    \item Stop\&Go: the vehicle is approaching the traffic light when red, so it stops and starts again when the phase changes into green. The algorithm suggests a prior deceleration to avoid the stop.
    \item Last second braking: in this scenario the vehicle is approaching the traffic light when green, but the phase will change when the vehicle is close to the semaphore and an acceleration is unfeasible, hence a hard braking is necessary. 
    Here the warning system informs the driver to reduce speed in advance so that a hard braking is avoided.
    \item Unnecessary stop: here the vehicle is approaching the traffic light when green, but the phase will change shortly. The STLA tells the driver to accelerate (respecting road speed limit and guaranteeing vehicle safety) so that the vehicle can pass when the light is still green.
\end{itemize}
Moreover, the MTLA is able to advise the driver on how to take a green wave. Two possible levels of guidance are issued to the driver:
\begin{enumerate}
    \item Green Warning: it encourages the driver to increase the speed in order to take one or more green TLs without stopping.
    \item Red Warning: it suggest to the driver to reduce speed.
\end{enumerate}
It is noteworthy that if no modification of the speed is requested to the driver, no warning is issued.\newline
In \cref{TLA_scheme}, a scheme of the TLA system is presented. Inputs are reported on the left side and are classified as static, dynamic or both static and dynamic. First step is the abscissa computation (``Localization" block), which is then used in the ``Activation Check" block to trigger the computation of reference acceleration and warning (``Reference Generation" \& ``Warning Definition" block). 

\subsection{Hypotheses}
The assumptions that are used in the development of the algorithms are the following:
\begin{enumerate}
    \item A point mass model is adopted to model the vehicle, as only the longitudinal behavior is relevant. The speed $v$ is always tangent to the path and $v=\dot{s}$, and the acceleration is $a=\ddot{s}$. Note that the reference point for the point mass model is defined with respect to the front of the vehicle; this choice is motivated by the fact that when the vehicle needs to stop, the front should be before the stop line of the traffic light. It may be pointed out that in case of passing at the end of the green phase, it is the back that should be considered, but there is still the yellow phase that enables the vehicle to drive through the intersection safely.
    \item The abscissa of the traffic light is the value of the abscissa of the stop line.
    \item No yellow phase is considered, it is part of the red one. This could be adjusted by placing all (or part) of the duration of the yellow phase in the green one.
    \item The warning system is triggered when the vehicle is within 500 m of one or more traffic lights.

    \item No further vehicles, other than the ADAS equipped one, are present on the road.
\end{enumerate}

\subsection{Testing Scenario}

\begin{figure} 
    \centering
      \subfloat[\label{MTLW_TestRoad}]{%
         \includegraphics[keepaspectratio=true,scale=0.5]{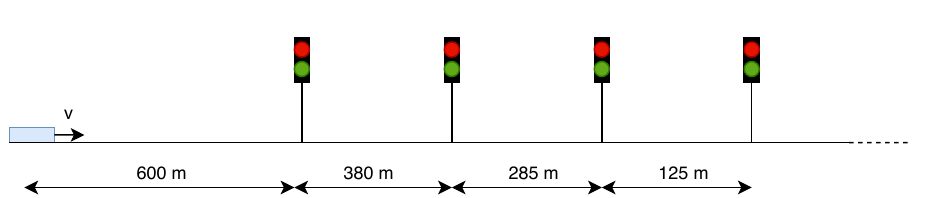}}\\
  \subfloat[\label{TLreal_pos}]{%
         \includegraphics[keepaspectratio=true,scale=0.3]{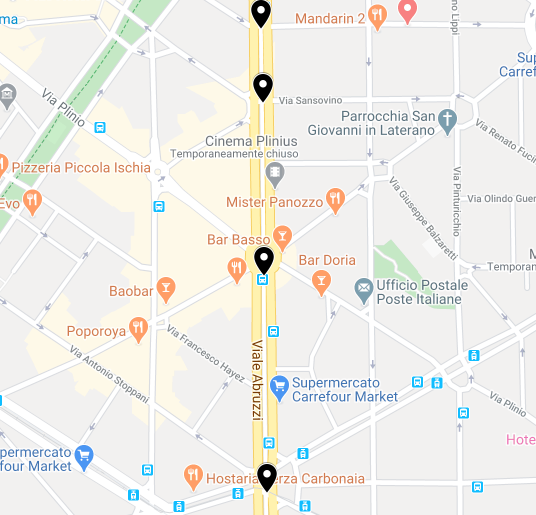}}%
    \hfill
  \subfloat[\label{TLreal_phase}]{%
         \includegraphics[keepaspectratio=true,scale=0.3]{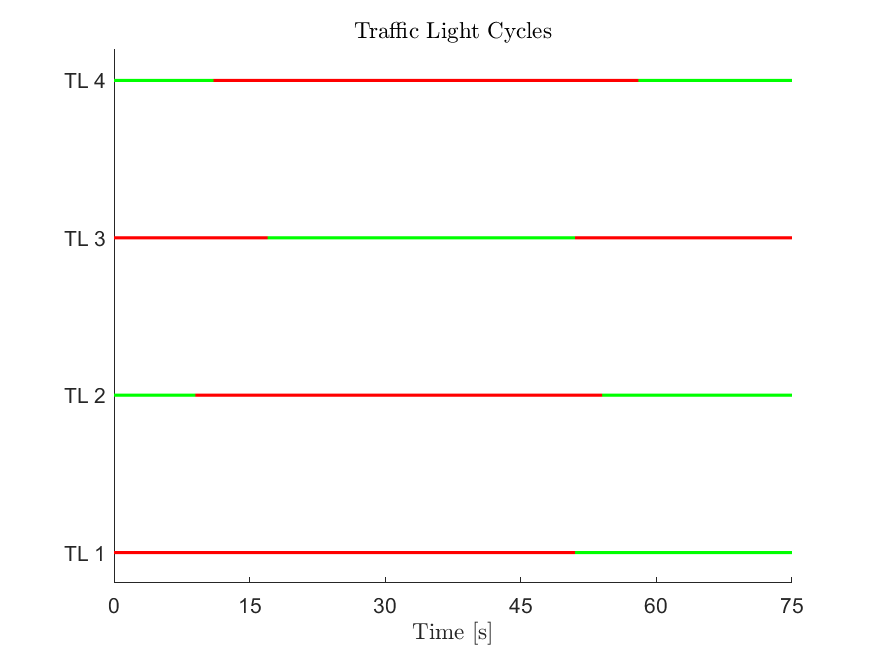}}
  \caption{(a) Sketch of test road showing the locations of the traffic lights, (b) Traffic lights positions on map, (c) Phase cycles of the four traffic lights }
\end{figure}

The CarMaker-Simulink environment is used to perform the testing. The main goal is to demonstrate that the MTLA system is able to help the driver in taking a green wave. To do so, a driver that receives no warning (driver 1) is compared with a driver that receives guidance from the GLOSA systems and follows it perfectly (driver 2). The initial speed is kept constant for both drivers until action is required to deal with traffic lights. \newline
The test road scenario described in \cref{MTLW_TestRoad} is a $1500~\text{m}$ long straight road, where four traffic lights with a $75~\text{s}$ cycle are placed. Traffic light location, phases and timing are based on real TLs located in the  city of Milan \cite{AMAT}. Their position and relative phases are reported respectively in \cref{TLreal_pos,TLreal_phase}.\newline
It is important to note that communication (or connectivity) is simulated in the CarMaker-Simulink simulation environment; however, it takes into consideration latency as well as lost packets that are associated with 5G. It is worth mentioning that the active map itself is simulated in a way that does not hinder the primary purpose of the simulation-based testing, and latency and reliability of communication (10 ms and 99.9\% respectively) are similar to what is reported in \cite{khayyat2021development}.


\section{Non-optimal MTLA}
The MTLA aim is to calculate a reference acceleration profile and to issue a warning that enables the vehicle to pass as many green traffic light as possible, based on all the inputs mentioned before. The warning suggests modifications to the velocity according to a reference acceleration profile.\newline
The four TLs ahead of the vehicle are iteratively analyzed every time the algorithm is triggered and the possibility to get one or more greens is subjected to the following conditions:
\begin{itemize}
    \item a reference velocity  that respects road limits and allows to pass at green phase has to be found for each TL;
    \item if an acceleration/deceleration maneuver is required to the driver, its safety, comfort and feasibility need to be checked;
    \item a common reference velocity  among the considered traffic lights needs to be found.
\end{itemize}
The above mentioned velocity and acceleration are selected respectively from a velocity and an acceleration range. Indeed, for each traffic light a minimum and maximum velocity that permit the driver to get a green are defined, together with the corresponding accelerations. Through a comparison between the computed reference velocity with the actual vehicle velocity, the warning is issued.\newline
In order to calculate the velocity and acceleration reference profiles, a certain type of motion needs to be chosen. For instance, in literature the velocity range is computed considering a constant speed profile, then the maximum velocity from this interval is considered as the reference one \cite{ MPC_MTL1, MPC_MTL2}. The drawback of the constant velocity profile is that the time needed for the driver to reach the reference velocity is not taken into account; hence, it is not possible to use it directly as reference for the warning. In fact, in \cite{ MPC_MTL1, MPC_MTL2}, the reference speed is used in a cost function which is then optimized considering the constraint of vehicle dynamics to calculate the control action.\newline
In this work,  the driver acceleration phase is considered in the computation of the velocity range: it is represented through a uniformly accelerated motion. This profile approximates the vehicle behavior, but it has been chosen as a trade-off between accuracy and simplicity.\newline
The computation of the vehicle velocity range required to reach a green traffic light is here analyzed. The following hypotheses are assumed:
\begin{enumerate}
    \item if the phase of the traffic light under analysis is green, the feasibility to get either the actual or the following green phase is analyzed;
    \item if the phase of the traffic light under analysis is red, the feasibility to get only the first green is analyzed;
    \item if the first traffic light in front of the vehicle is analyzed, a uniformly accelerated motion (UAM) until the traffic light is supposed;
    \item if a traffic light different from the first one is analyzed, the following  motion profile is assumed: a uniformly accelerated motion up to the first semaphore is considered, then a constant speed motion (CSM) is adopted (UAM+CSM).
\end{enumerate}
From here on, in order to differentiate among UAM and UAM+CSM, the first traffic light is referred to as first traffic light, while the next ones as $i^{th}$ traffic lights.\newline
The motion profiles described in hypothesis 3 and 4 and the corresponding velocity range computation are analyzed in the next section. 
\subsection{First Traffic Light}

\begin{figure}[t]
    \centering
  \subfloat[\label{TL11_distance}]{%
 	\includegraphics[keepaspectratio=true,scale=0.8]{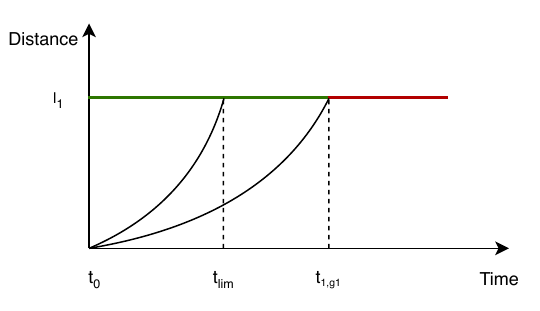}}\\[-1ex]%
  \subfloat[\label{TL1_distance_red}]{%
         \includegraphics[keepaspectratio=true,scale=0.8]{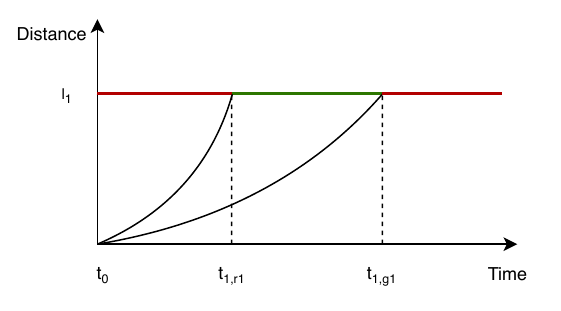}}
  \caption{(a) Uniformly accelerated motion to reach the first traffic light during its first green phase when the green is on. (b) Uniformly accelerated motion to reach the first traffic light during its first green phase when the red is on. $t_0$ is the algorithm triggering time.}
\end{figure}

The UAM in \cref{MUA} is adopted to generate the speed interval for the first traffic light:
\begin{equation}
\label{MUA}
\begin{cases}
d = vt +\frac{1}{2}at^2\\
v_\text{target} = v +at
\end{cases}
\end{equation} 
where; $v$ is the actual speed, $v_\text{target}$ is the target speed reached after the acceleration phase, $d$ is the acceleration distance which is set equal to the distance between the front of the vehicle and the traffic light ($d =l_1$).\newline
It is important to note that \cref{MUA} is a set of two equations with three unknowns: $v_\text{target}$, $a$ and $t$, so a parameter still need to be fixed in order to obtain a unique solution. To do so, two cases  can be distinguished according to the phase of the traffic light: green or red.
\paragraph{Green Phase}
If the first green of the TL is taken into consideration, the minimum velocity is the one that allows crossing the traffic light when the phase is just starting to shift to red. The time at which the phase shift occurs is the remaining time of the first green phase ($t_{1,g1}$), where the first index refers to the number of the TL and the second to the number of the green phase. By substituting $t=t_{1,g1}$ in \cref{MUA}, minimum velocity and acceleration can be computed. The maximum velocity is the one that allows getting the green state in the shortest time possible. So, in order to respect the rules of the road, the maximum velocity is  set equal to $v_t=v_\text{lim,road}$. By substituting $v_t=v_\text{lim,road}$ in \cref{MUA}, maximum velocity and acceleration  can be computed too. An example of the two profiles is represented in \cref{TL11_distance}. $t_\text{lim}$ is the time needed to reach the traffic light with speed equal to $v_\text{lim,road}$, i.e. the minimum possible time assuming a UAM to the first traffic light.
If the second green phase is under analysis the minimum time needed to reach the traffic light correspond to the time of the red to green phase shift $t_{1,r1}$. The maximum one coincides with the time of the second green to red phase shift $t_{1,g2}$. In this way the possibility to get either the first or the second red phase is avoided. By substituting $t_{1,g2}$ and $t_{1,r1}$ in \cref{MUA}, the minimum and maximum velocity and acceleration profiles can be computed.
\paragraph{Red Phase}
The minimum time needed for the driver to reach the first available green phase is the one of the red to green phase shift $t_{1,r1}$. The maximum one corresponds to the end of the green phase and so to the time of the green to red phase shift $t_{1,g1}$. The profiles are shown in \cref{TL1_distance_red}. By substituting the previously mentioned times $t_{1,g1}$ and $t_{1,r1}$ in \cref{MUA}, the final formulation can be obtained.
\subsection{$\text{i}^\text{th}$ Traffic Light}
\begin{figure}[t]
	\centering
	\includegraphics[keepaspectratio=true,scale=0.8]{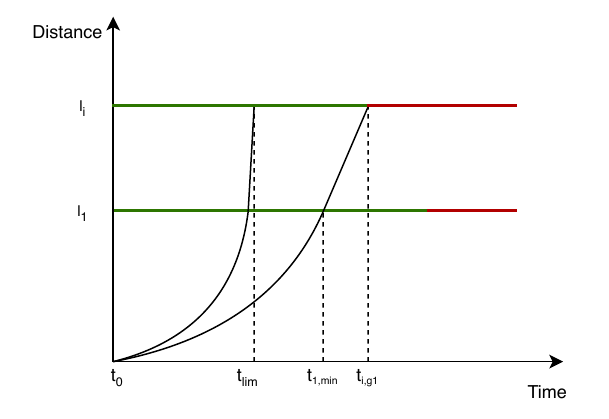}
	\caption{Motion profile to reach the $i^{th}$ traffic light during its first green phase when the green is on.}
	\label{MUACi1}
\end{figure}

\begin{figure}[t]
	\centering
	\includegraphics[keepaspectratio=true,scale=0.8]{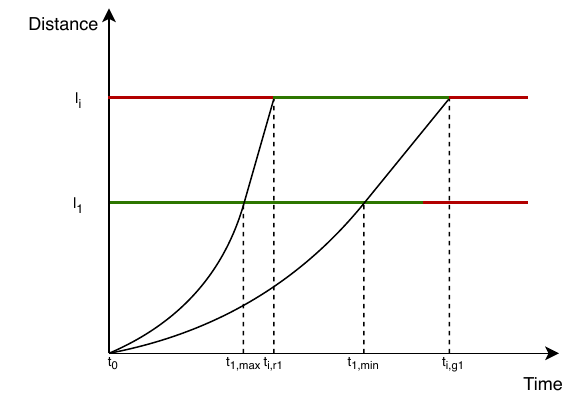}
	\caption{Motion profile to reach the $i^{th}$ traffic light during its first green phase when the red is on.}
	\label{TLij_red}
\end{figure}

A generic traffic light after the first one is analyzed. The motion profile is based upon an acceleration up to the first traffic light, followed by a constant velocity motion up to the $i^{th}$ traffic light. Such motion profile is described by \cref{MUAC_eq}.
\begin{equation} 
\label{MUAC_eq}
\begin{cases}
d_1 = vt_1 +\frac{1}{2}at_1^2\\
v_t = v +at_1 \\
d_2 = v_t t_2
\end{cases}
\end{equation}
where: $v$ is the actual speed of the vehicle, $v_\text{target}$ is the target speed reached after the acceleration phase and $d_1$ is the distance traveled during the acceleration phase, i.e. the distance $l_1$ between the vehicle and the first traffic light. Then, $d_2$ is the space driven at constant speed, equal to the distance between the $i^{th}$ traffic light and the first one, and $t_2$ is the difference between the duration of the overall maneuver $t_{tot}$  and that of acceleration phase $t_1$.\newline
Note that \cref{MUAC_eq} is a set of three equations with four unknowns: $t_1$, $a$, $v_t$, $t_\text{tot}$. In order to find a solution a parameter still needs to be fixed. Again, two cases need to be distinguished according to the actual traffic light phase: green or red.
\paragraph{Green Phase}
No constraints on the minimum time to reach the traffic light during its first green exists, thus the maximum velocity is set equal to the road limit speed. By imposing $v_\text{target}=v_\text{lim,road}$ in \cref{MUAC_eq},  maximum velocity and acceleration are computed. The maximum time to get the first green of the $i^{th}$ TL is the time of the green to red phase change of that TL. Given the remaining time of the green phase $t_{i,g1}$, the total maneuver time is equal to the latter. By imposing $t_\text{tot} = t_{i,g1}$ in \cref{MUAC_eq} minimum velocity and acceleration can be computed. The solution is in \cref{MUAC_i1min}:
\begin{equation}
\label{MUAC_i1min}
\begin{cases}
a_{\min}=2\dfrac{l_1}{t_{1,\min}^2}-2\dfrac{v}{t_{1,\min}} \\[4ex]
v_{\min}=\dfrac{\splitfrac{l_i+l_1-t_{i,g1}v}{+\sqrt{(-l_i-l_1 +t_{i,g1}v)^2+4t_{i,g1}v(l_i-l_1)} }}{2t_{i,g1}}\\[1ex]
t_{1,\min}=t_{i,g1}-\dfrac{l_i-l_1}{v_{\min}}
\end{cases}
\end{equation}
The profiles obtained with the minimum and maximum velocities are shown in \cref{MUACi1}, respectively on the right and left.
\paragraph{Red Phase}
The minimum time needed for the driver to reach the first green phase of the $i^{th}$ TL is the one of the red to green phase shift $t_{i,r1}$. The maximum one is the time of the end of the green phase $t_{i,g1}$. This scenario is represented in \cref{TLij_red}, respectively on the left (minimum time) and right (maximum time). 

\subsection{MTLA Algorithm Steps}
\begin{figure}[htb!]
	\centering
	\includegraphics[keepaspectratio=true,scale=0.6]{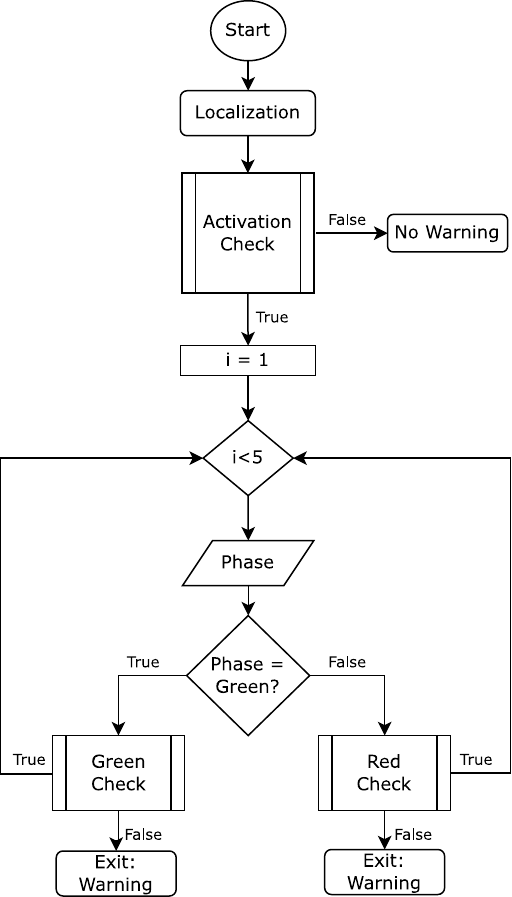}
	\caption{Non-Optimal Multiple Traffic Light Advisor algorithm flow chart. Index $i$ represents the number of the traffic light in analysis.}
	\label{MTL_FC}
\end{figure}

\begin{figure}[htb!]
	\centering
	\includegraphics[keepaspectratio=true,scale=0.6]{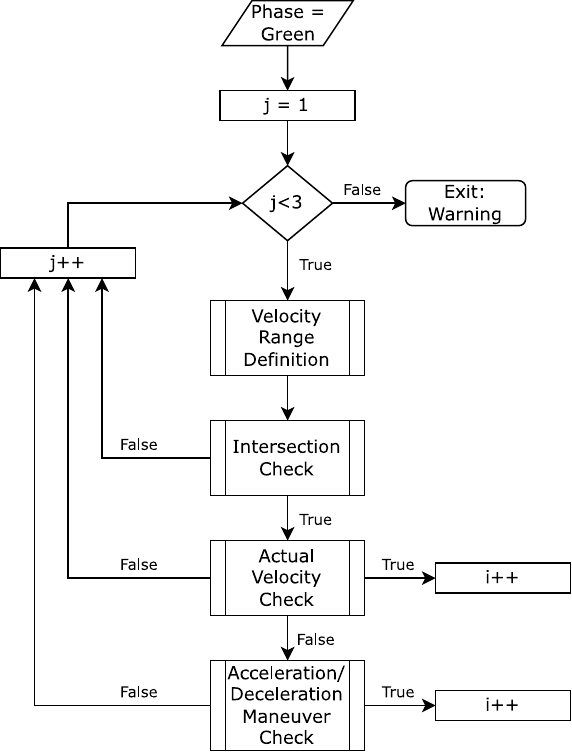}
	\caption{Non-Optimal Multiple Traffic Light Advisor algorithm flow chart - Green Check. Index $j$ refers to the green phase analyzed. }
	\label{MTL_FC_Green}
\end{figure}

\begin{figure}[t] 
	\centering
	\includegraphics[keepaspectratio=true,scale=0.6]{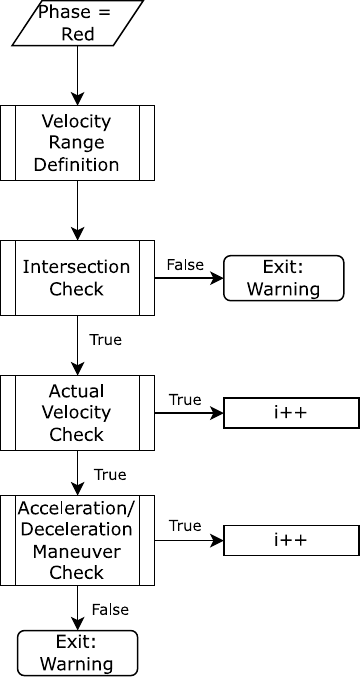}
	\caption{Non-Optimal Multiple Traffic Light Advisor algorithm flow chart - Red Check}
	\label{MTL_FC_Red}
\end{figure}

After defining the speed profile for each TL, the working principle of the algorithm which finds the final profile is shown in the flow chart of \cref{MTL_FC}, the ``Green Check" and ``Red Check" blocks are further illustrated in  \cref{MTL_FC_Green,MTL_FC_Red} respectively. Index $i$ represents the number of the traffic light in analysis, while index $j$ refers to the green phase analyzed. The maximum value of $j$ has been set to $2$ in order not to perform too many iterations each time the algorithm is run. If $j=3$ it would mean to pass the third green phase, in case the actual phase is green, or the second green, in case it is red, and to do so such a low speed would be required that it would result disturbing for the driver.
The first two steps from \cref{MTL_FC} are the Localization and the Activation Check. They calculate the abscissa of the vehicle and check the presence of a TL within a certain distance, called horizon. Once the algorithm finds the first of the four traffic lights within the horizon, the system should be always active until the last semaphore is passed. If the distance between two subsequent TLs is greater than the horizon, the system is deactivated. To avoid so, the horizon for this application is set to $500 \ \text{m}$. This value has been decided considering distances between semaphores in \cite{AMAT}. Once the algorithm is activated, an iterative cycle which analyzes the four traffic lights ahead of the vehicle starts.\newline
The feasibility to get a green phase is examined and whenever a feasible maneuver to reach the $i^{th}$ traffic light is found, the following variables are stored inside the algorithm: reference velocity ($v_\text{ref}$), reference acceleration ($a_\text{ref})$, warning to be issued and interval of admissible velocities ($v_{\text{adm},i}=[v_{\min,i},v_{\max,i}]$).\newline
If it is not possible to pass the traffic light under examination during a green phase, the algorithm terminates and the last stored warning is issued.
The ``Green Check" block, (\cref{MTL_FC_Green}) iterates according to the index $j$ and analyzes the possibility to get either the first ($j=1$) or second ($j=2$) green phase of the $i^{th}$ traffic light. If neither of them can be get, the algorithm terminates and a warning is issued to the driver. The first step of the Green Check is the Velocity Range Definition, this step computes the required velocity range to get the actual green phase $v_\text{req,i}$ according to the indexes i and j as previously explained. The second step is the Intersection Check. In this block the required velocity range to get the $i^{th}$ green with the one needed to get the $i-1^{th}$ green are intersected as in \cref{VminInters}. The aim is to find a velocity range that allows to pass both the $i^{th}$ and $i-1^{th}$ traffic lights. 
\begin{equation}
    v_{\text{adm},i} = v_{\text{req},i} \cap v_{\text{adm},i-1}
    \label{VminInters}
\end{equation}
If the first traffic light is studied ($i=1$), the admissible velocity range is defined by the road minimum ($20~\text{km}\text{h}^{-1}$) and maximum ($50~\text{km}\text{h}^{-1}$) admissible velocities. On the other hand, when $i>1$ the admissible interval is already defined from the $i-1^{th}$ iteration. Now, if \cref{VminInters} results in an empty interval, it is not possible to reach the $i^{th}$ traffic light when its $j^{th}$ green phase is on, so the possibility to get the next one is then analyzed ($j+1$). On the contrary if the interval is non-empty, $v_{\text{adm},i}$ is defined.\newline
Then, the ``Actual Velocity Check" (AVC) block analyzes the possibility for the driver to keep their current speed rather than performing an acceleration or deceleration maneuver. If the last check is satisfied, the next traffic light is analyzed ($i+1$), if not, the ``Acceleration/Deceleration Maneuver" Check (AMC) is performed. Given that the diver cannot keep its actual speed constant, the safeness of the acceleration or deceleration maneuver is checked. If the AMC block result is positive the driver can get the $j^{th}$ green of the $i^{th}$ traffic light. The index $i$ is increased by one unit and the next traffic light is studied. If the result is negative, the next green phase is analyzed.\newline
The steps performed by the ``Red Check" block are the same as for the green case. The only difference is that, if the actual phase is red, only the possibility to get the first green is evaluated \cref{MTL_FC_Red}.

\section{Optimal MTLA}
\begin{figure}[t]
	\centering 
	\includegraphics[keepaspectratio=true, width = 0.5 \textwidth]{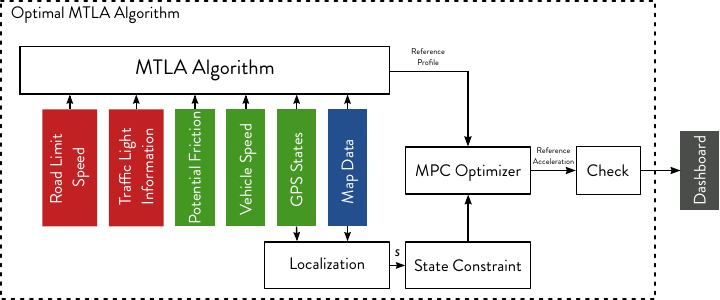}
	\caption{Optimal Traffic Light Advisor algorithm scheme}
	\label{OMTLA_scheme}
\end{figure}

The reference acceleration profile calculated by the non-optimal MTLA is optimized by means of MPC and then the new optimal acceleration profile is used to trigger the warning. An overview of the system is presented in \cref{OMTLA_scheme}, the profile generated by the non-optimal MTLA algorithm (``Non-Optimal MTLA Algorithm" block) is fed as reference to the model predictive controller (MPC block) along with state constraints (computed in the ``State Constraints" block).%
With respect to similar works in literature, some noteworthy differences include the following: 
\begin{enumerate}
    \item The cost function here proposed not only includes the error with respect to a reference profile but also penalizes the jerk of the vehicle: in this way it is possible to increase driving comfort \cite{MPC_jerk};
    \item a  UAM+CSM reference profile is adopted rather than only a CSM one \cite{MPC_MTL1};
    \item the dynamic variability of traffic lights phases makes the solution space of the optimal problem non-convex and a constant speed reference motion profile is used in literature to tackle this problem. Here, through the use of a more accurate profile (UAM+CSM), a solution can be found faster and it is more likely to be a global-optimum\cite{MPC_MTL1};
\end{enumerate}

Since literature lacks detailed explanation of how to consider TL phases into state constraints, an algorithm for generating the position constraints according to TL color is proposed and discussed.

\subsection{Motivation}
The algorithm in \cref{MTL_FC} generates a UAM+CSM profile, which could be use as reference for defining the warning; however, it has some weaknesses. 
The UAM+CSM reference profile is not continuous in acceleration and it does not consider vehicle dynamics, for these reasons it may be difficult for the driver to follow  the suggested maneuver or it may lead to discomfort due to high jerk values. In order to overcome the above mentioned problems, the reference profile is optimized through an MPC approach.
\subsection{Problem Formulation}
\begin{figure}[t]
    \centering
        {\includegraphics[keepaspectratio=true,scale=0.35]{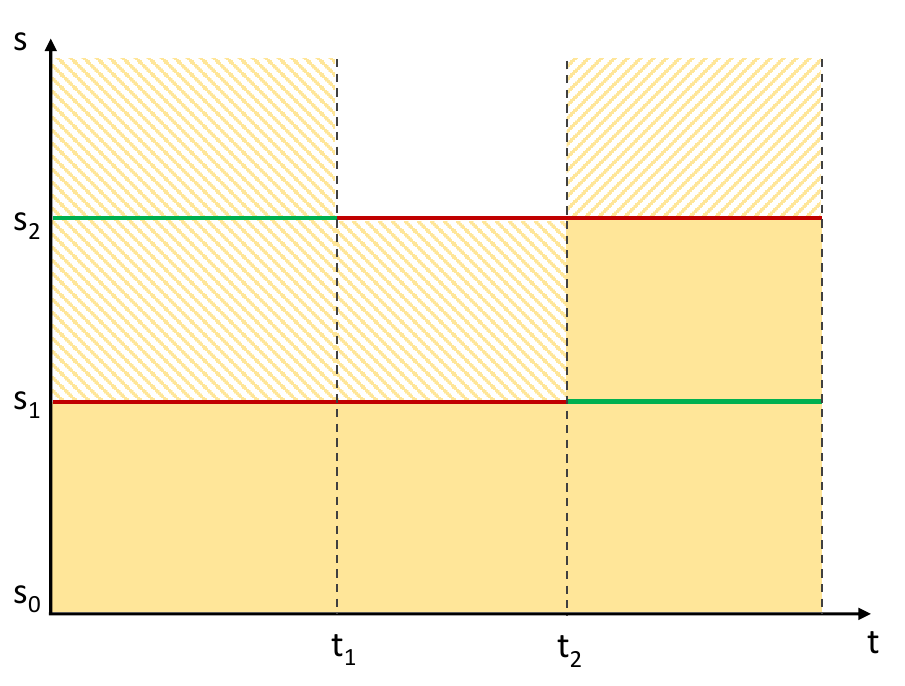}}
\caption{Graphical explanation of the algorithm for state constraints formulation. $s_0$ is the abscissa of the vehicle on the path, $s_1$ and $s_2$ are the abscissa of the first and second traffic light respectively, the lines color represents the phase (green or red), while $t_1$ and $t_2$ are the times of the phase change for the first and second TLs}
\label{fig:region_ex}
\end{figure}

\begin{figure}[htb!]
    \centering
     \includegraphics[keepaspectratio=true,scale=0.65]{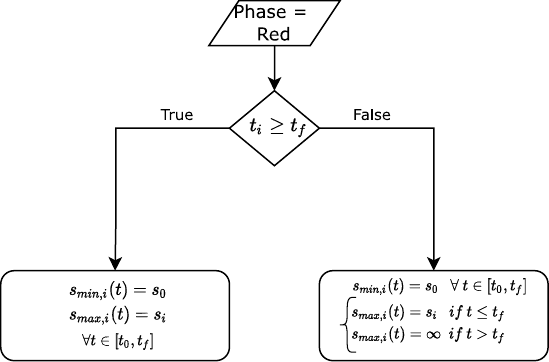}
\caption{Red Check block - flow chart of the algorithm for the definition of the $i^{th}$ traffic light state constraints}
\label{FlowChart_SC_3}
\end{figure}

\begin{figure}[htb!]
    \centering
     \includegraphics[keepaspectratio=true,scale=0.65]{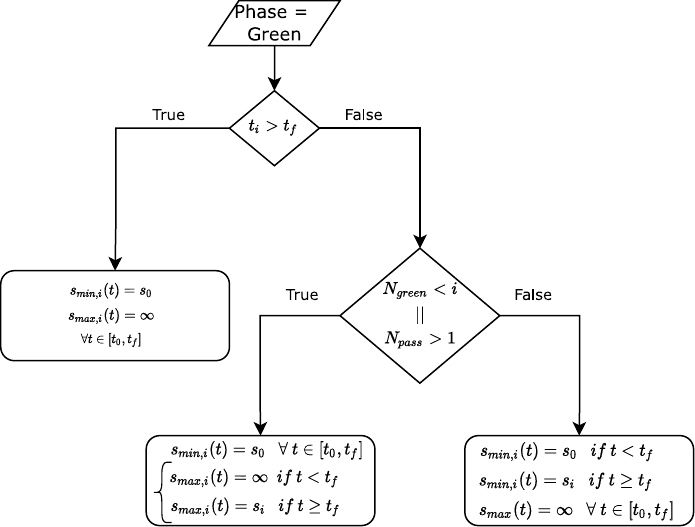}
\caption{Green Check block - flow chart of the algorithm for the definition of the $i^{th}$ traffic light state constraints}
\label{FlowChart_SC_4}
\end{figure}

The OCP in \cref{ProbFormulation} to be solved is defined by the cost function, dynamic model and constraints. To obtain a smooth acceleration profile the problem has been defined so to minimize the predicted jerk while still following the reference trajectory. Concerning the dynamic model, a simplified one that is suitable for our applications is considered; a one degree of freedom plane model is used, thus neglecting road inclination and lateral dynamics. The input  constraints limit the net maximum braking and net traction force. Moreover, state constraints are used to avoid passage of the vehicle with red traffic light. To do so the vehicle position is limited to some region of the time-space plane as shown in \cref{fig:region_ex}.\newline
It is noteworthy that only two traffic lights and one phase shift for each of them are considered. This is due to the fact that, having chosen the length of the prediction horizon $t_f$ equal to $6~\text{s}$, more than two TLs cannot be encountered in this interval and more than one phase shift cannot occur. The length of the prediction horizon is chosen as a trade off between small enough discretization and computational effort while still guaranteeing good optimization results.
\begin{figure}[t]
    \centering
     \includegraphics[keepaspectratio=true,scale=0.6]{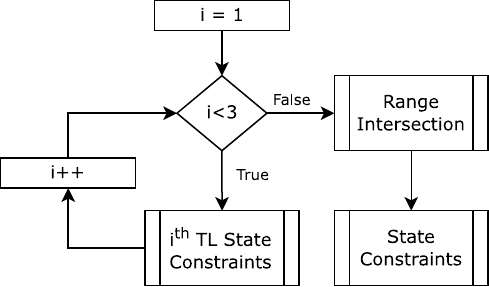}
\caption{Flow chart of the algorithm for the definition of state constraints}
\label{FlowChart_SC_1}
\end{figure}
\begin{figure}[t]
    \centering
     \includegraphics[keepaspectratio=true,scale=0.6]{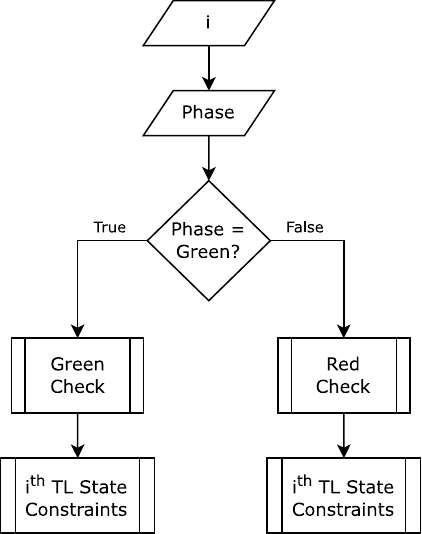}
\caption{$i^{th}$ TL State Constraints block - flow chart of the algorithm for the definition of the $i^{th}$ traffic light state constraints}
\label{FlowChart_SC_2}
\end{figure}\newline
In order to calculate the admissible region in \cref{fig:region_ex}, the algorithm represented by the flow charts in \cref{FlowChart_SC_1,FlowChart_SC_2} is used. The flow charts of the ``Red Check" block is shown in \cref{FlowChart_SC_3}. If the time of the phase shift is larger or equal than the prediction horizon, the vehicle should be ahead of the TL for the whole horizon. Otherwise it can be beyond the TL only after the phase shift. The flow chart of the ``Green Check" block is illustrated in \cref{FlowChart_SC_4}. If the time of the phase shift is larger than the prediction horizon, the vehicle can be ahead or beyond the TL. Contrary, if the time is lower or equal, a further analysis is performed on the possibility to pass the TL during the actual green phase. This is done by checking if the MTLA algorithm calculates that it is not possible to pass the $i^{th}$ traffic light ($N_\text{green}<i$) or to pass it during the second green phase ($N_\text{pass}>1$). If one of these two conditions is verified the vehicle should be ahead of the TL after the phase shift, otherwise it should be beyond it.
Once the admissible regions are obtained for the two TLs, the ``Range Intersection" block (\cref{FlowChart_SC_1}) is executed. It consists of making the intersection of the state constraints of the single TLs as in \cref{constr_inter}:
\begin{equation}
\small
    \label{constr_inter}
    [s_{\min}(t),s_{\max}(t)] = [s_{\min,1}(t),s_{\max,1}(t)] \cap [s_{\min,2}(t),s_{\max,2}(t)]
\end{equation}
In addition to the position state constraint, also velocity, acceleration and jerk are constrained. Hence, the OCP to be solved is the following:
\begin{equation}
\small 
\left\{
\begin{split}
&\underset{v(t), a(t), j(t)}{\min} \int\limits_{0}^{t_f} w_v (v(t)-v_\text{ref}(t))^2 +w_a (a(t)- a_\text{ref}(t))^2  \\[-2ex]
& \quad \quad \quad \quad \quad \quad \quad +w_j j(t)^2 \text{d}t \\[1ex]
&\text{s.t } \quad  \begin{array}[t]{ll} \text{m} a(t) + \frac{1}{2}A_f \rho C_d v^2(t) + \text{m} \text{g} C_r = F(t) &\quad \forall t \in [0,t_f]\\
j(t) = \dot{a}(t) &\quad \forall t \in [0,t_f]\\
F_{\min} \le F(t) \le F_{\max}&\quad \forall t \in [0,t_f]\\
s_{\min}(t) \le s(t) \le s_{\max}(t) &\quad \forall t \in [0,t_f]\\
v(t) \ge 0  &\quad \forall t \in [0,t_f]\\
a(t) \le a_{\max}  &\quad  \forall t \in [0,t_f]\\
j_{\min} \le j(t) \le j_{\max} &\quad \forall t \in [0,t_f]
\end{array}
\end{split}
\right.
\label{ProbFormulation}
\end{equation}
Where $j(t)$ is the jerk, $a_\text{ref}(t)$ and $v_\text{ref}(t)$ are the reference acceleration and velocity from the non-optimal MTLA algorithm, and $w_a$, $w_v$ and $w_j$ are the weights of acceleration, velocity and jerk cost terms respectively. $A_\text{f}$ is the frontal area of the vehicle, $\rho$ is the air density, $C_\text{d}$ is the aerodynamic drag coefficient, $m$ is the vehicle mass, $C_\text{r}$ is the rolling resistance coefficient and $F$ is the generalized longitudinal tire force. $F_{\min} (<0)$ and $F_{\max} (>0)$ represent the maximum braking and traction forces, respectively.
\section{Simulation}
In this section, results of test cases are shown. Position, speed, acceleration and energy consumption are considered. Moreover, to prove that the developed MTLA is able to effectively reduce vehicle consumption, a simple energetic analysis is proposed considering instantaneous and average energy consumption.\newline
Instantaneous energy consumption is computed starting from the instantaneous power $P$ ($P=T\omega$ with $T$ being the motor torque and $\omega$ the motor angular velocity) as in \cref{IEC}:
\begin{equation}
    I.E.C.=P\frac{\text{travelled time}}{\text{travelled distance}} 
    \label{IEC}
\end{equation}
Average energy consumption is computed starting from the instantaneous one, an in \cref{AEC}:
\begin{equation}
\begin{aligned}
    A.E.C._n&=\frac{I.E.C._1+...+I.E.C._n}{n}\ \ 
    \\
    A.E.C._{n+1}&=\frac{I.E.C._{n+1}+...+nA.E.C._n}{n+1}\ \ 
        \label{AEC}
    \end{aligned}
    \end{equation}
\subsection{Non optimal MTLA}
\begin{figure}[t] 
	\centering
	\includegraphics[keepaspectratio=true,scale=0.5]{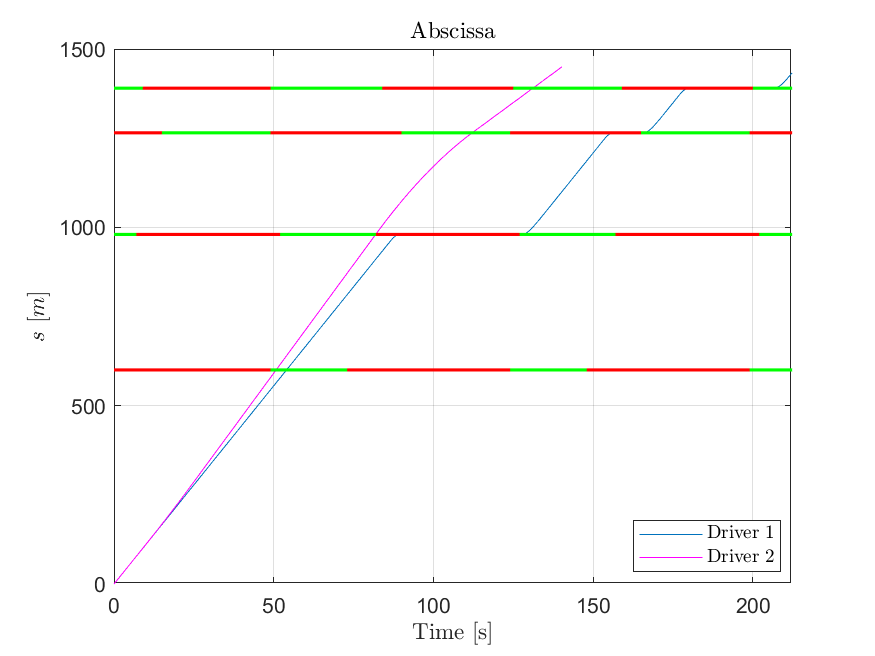}
	\caption{MTLA Test Case - abscissa of driver 1 and driver 2}
    \label{MTLWr_C1_s}
\end{figure}

\begin{figure}[t] 
	\centering
	\includegraphics[keepaspectratio=true,scale=0.5]{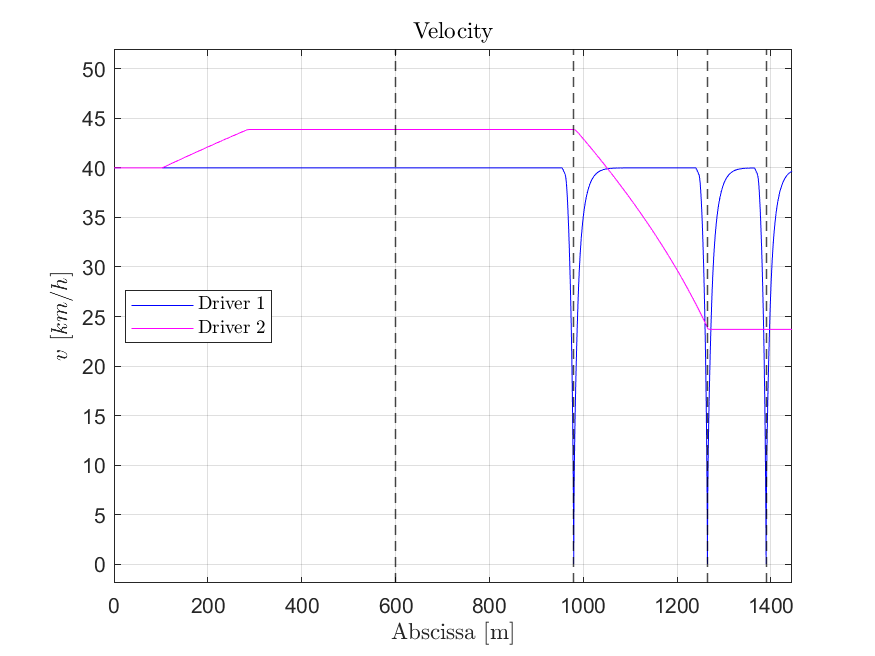}
	\caption{MTLA Test Case - velocity profiles of driver 1 and driver 2}
    \label{MTLWr_C1_v}
\end{figure}

\begin{figure}[t] 
	\centering
	\includegraphics[keepaspectratio=true,scale=0.5]{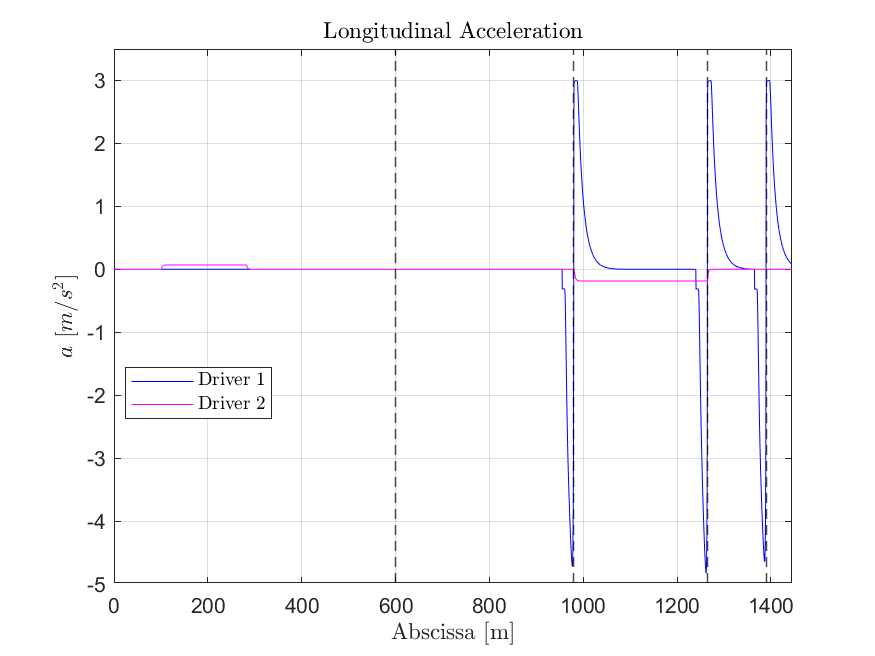}
	\caption{MTLA Test Case - acceleration profiles of driver 1 and driver 2}
    \label{MTLWr_C1_a}
\end{figure}

\begin{figure}[t]  
	\centering
	\includegraphics[keepaspectratio=true,scale=0.5]{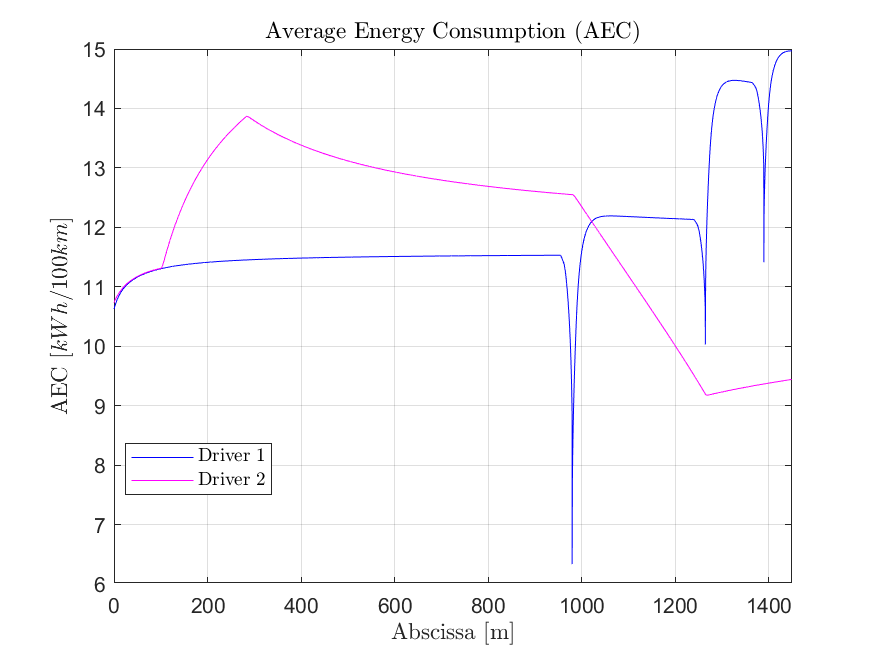}
	\caption{MTLA Test Case - Average Energy Consumption of driver 1 and driver 2}
    \label{MTLWr_C1_AEC}
\end{figure}

This test case highlights advantages of MTLA in terms of stops, travel time and vehicle consumption. From \cref{MTLWr_C1_s} it is visible how driver 1 is subject to three stops, thus increasing the total travel time to cross all the traffic lights. With the initial velocity of $40~\text{km}\text{h}^{-1}$ driver 1 is able to reach the first traffic light in its green phase, while it is not able to get the second one. On the contrary, driver 2 given an acceleration warning is able to avoid such stop. From \cref{MTLWr_C1_v,MTLWr_C1_a} it can be seen that driver 1 increases its velocity as soon as the first traffic light is in the horizon of activation of the algorithm. This acceleration allows also to get the third semaphore, but not the fourth. It is to be noticed that the vehicle stops accelerating before reaching the first traffic light (\cref{MTLWr_C1_v,MTLWr_C1_a}) because a velocity that allows reaching all the previously analyzed three green light is reached.\newline
When the second traffic light phase shifts to green again driver 1 accelerates and reaches the speed of $40~\text{km}\text{h}^{-1}$. While approaching the third traffic light, driver 1 starts decelerating to stop the vehicle again as it is red. The same deceleration/acceleration maneuver occurs while reaching the fourth traffic light. Given a deceleration warning, driver 2 is able to avoid these two stops. As soon as driver 2 overcomes the second traffic light, a new profile can be calculated which allows to pass also the fourth semaphore without stopping.\newline
It is noteworthy to to note that when the algorithm is triggered at first, no feasible solution could be found to reach the fourth traffic light; however, as the vehicle passes across the TLs, new profiles are calculated and a feasible UAM+CVM profile to cross the fourth semaphore is found. This is due to the algorithm's continuous update.
Moreover, it is visible from \cref{MTLWr_C1_a} that acceleration values of driver 2 are lower than driver 1, meaning higher maneuver comfort.
\cref{MTLWr_C1_AEC} shows the average energy consumption. It can be noticed how the final average consumption (after passing the four traffic lights) of driver 2 is lower than driver 1: $9.4~\text{kWh}\text{100~km}^{-1}$ for driver 2 versus $15~\text{kWh}\text{100~km}^{-1}$ for driver 1. Hence, considering these last values, MTLA shows a $37.3\%$  reduction in energy consumption.
\subsection{Optimal MTLA}
\begin{figure}[htb!]
	\centering
	\includegraphics[keepaspectratio=true,scale=0.5]{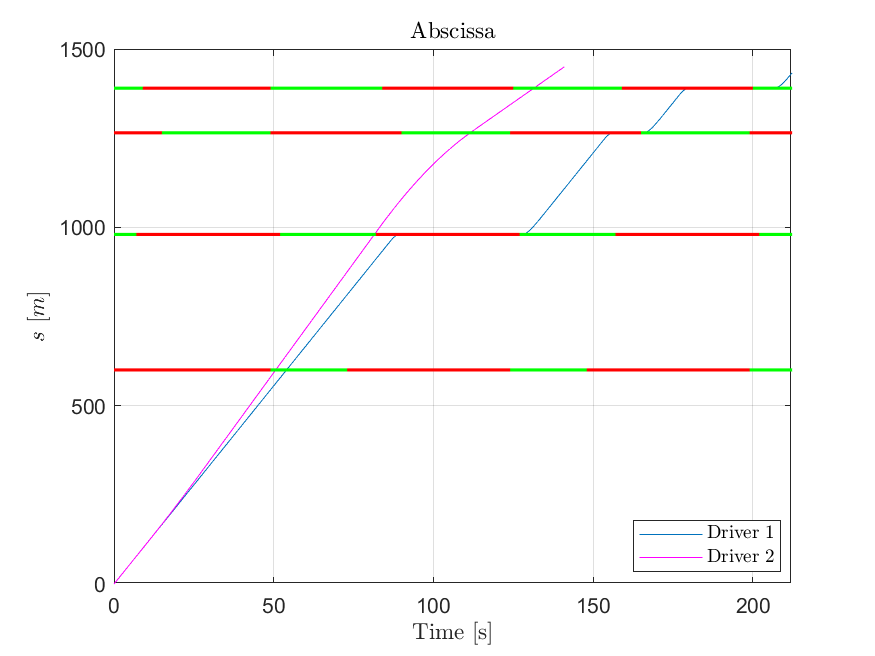}
	\caption{Optimal MTLA Test Case 1 - abscissa of driver 1 and driver 2}
   \label{Test_sV40}
\end{figure}

\begin{figure}[htb!]
	\centering
	\includegraphics[keepaspectratio=true,scale=0.5]{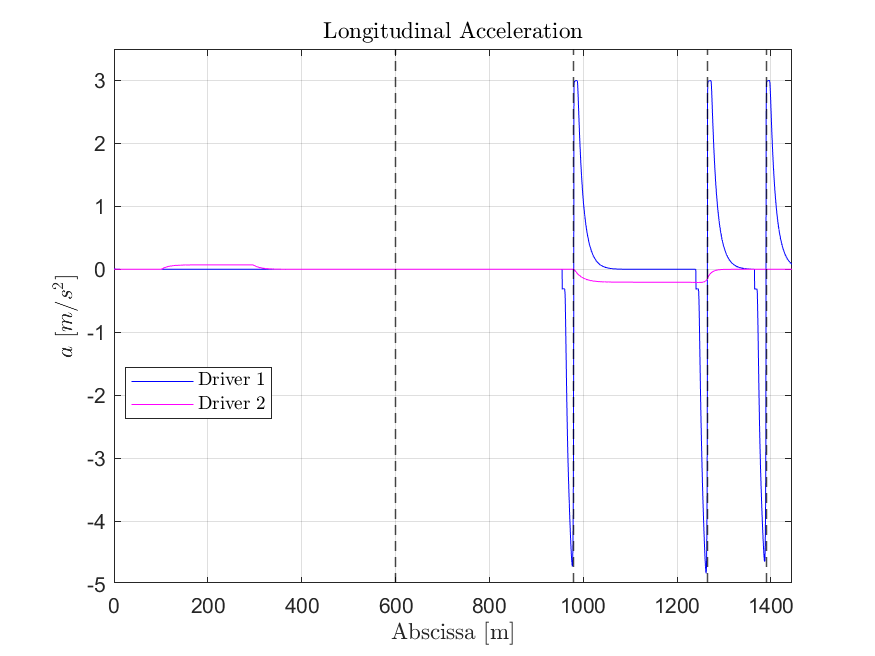}
	\caption{Optimal MTLA Test Case 1 - acceleration profiles of driver 1 and driver 2}
   \label{Test_aV40}
\end{figure}

\begin{figure}[htb!]
	\centering
	\includegraphics[keepaspectratio=true,scale=0.5]{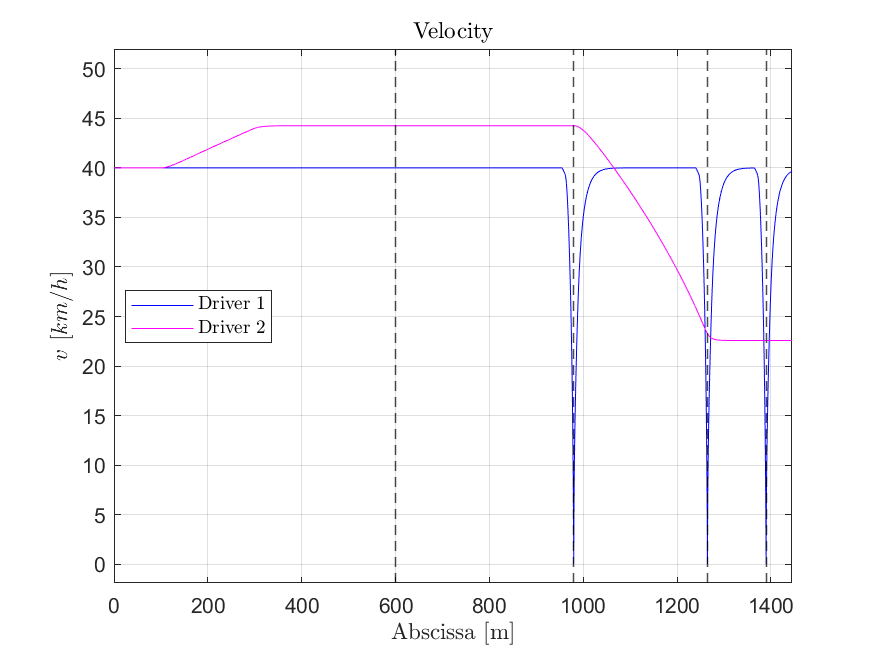}
	\caption{Optimal MTLA Test Case 1 - velocity profiles of driver 1 and driver 2}
   \label{Test_VelV40}
\end{figure}

\begin{figure}[htb!]
	\centering
	\includegraphics[keepaspectratio=true,scale=0.5]{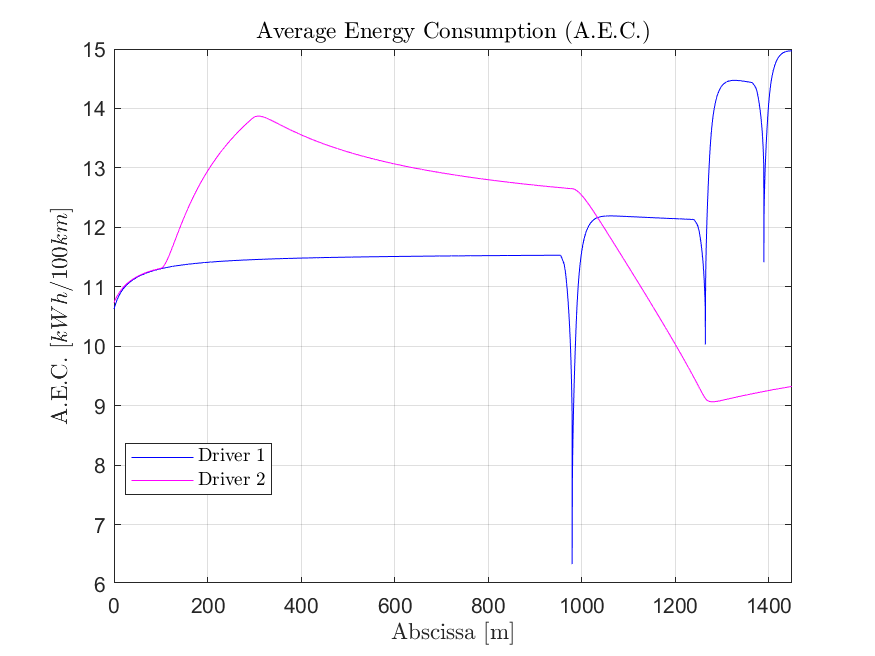}
	\caption{Optimal MTLA Test Case 1 - Average Energy Consumption of driver 1 and driver 2}
    \label{OMTLWr_C1_AEC}
\end{figure}

First, the comparison between the vehicle equipped with optimal MTLA system (driver 2)  and the one without any kind of warning system is reported (driver 1). Looking at \cref{Test_sV40} it can be noticed how driver 1 has to stop at three traffic lights, while driver 2, having followed the warning, is able to take a green wave and thus  reduce the total travel time. The comments done for the Non-Optimal MTLA are also valid  for this case since driver 1 is the same and driver 2 has the same behavior with the only difference that the acceleration profile is smoother. Acceleration and velocity of driver 1 and 2 are plotted with respect to abscissa in \cref{Test_aV40,Test_VelV40}. \cref{OMTLWr_C1_AEC} shows the average consumption (computed as in \cref{AEC}) for driver 1 and driver 2. As for the previous plots, comments done previously are  valid also for the energetic analysis. It can be appreciated how driver 1 final average consumption ($15~\text{kWh}\text{100~km}^{-1}$) is higher than  driver 2 ($9.3~\text{kWh}\text{100~km}^{-1}$), resulting in a $38\%$ reduction.
\begin{figure}[htb!]
	\centering
	\includegraphics[keepaspectratio=true,scale=0.5]{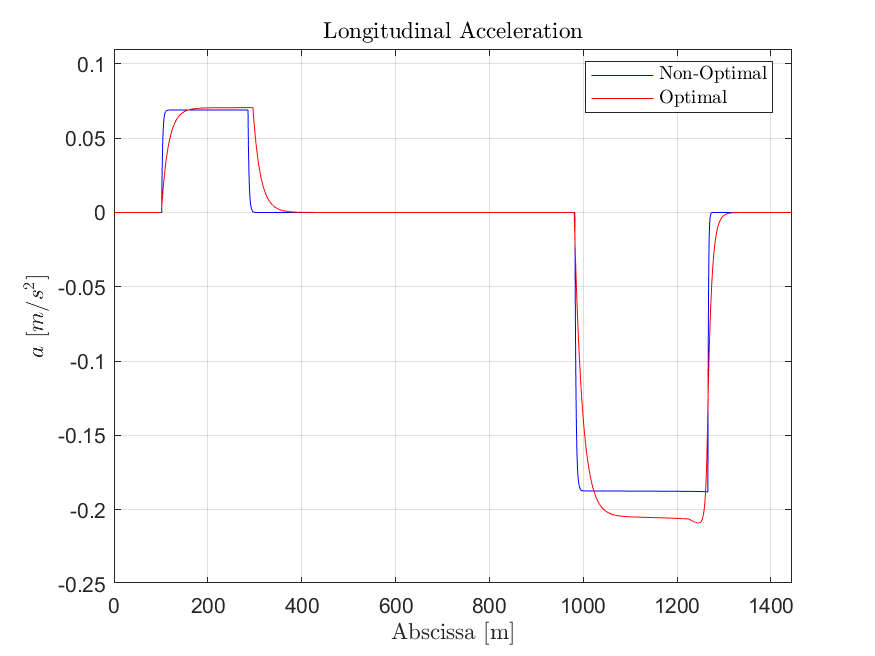}
	\caption{Optimal MTLA Test Case 1 - non-optimal and optimal acceleration profiles}
   \label{aV40}
\end{figure}
\begin{figure}[htb!]
	\centering
	\includegraphics[keepaspectratio=true,scale=0.5]{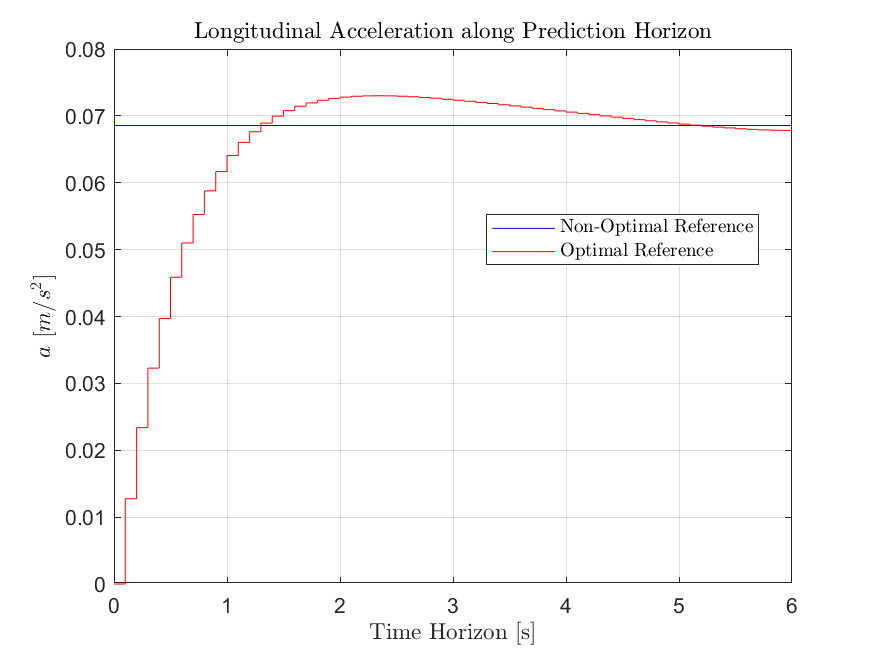}
	\caption{Optimal MTLA Test case 1 - non-optimal and optimal acceleration profiles along prediction horizon at given instant of simulation}
   \label{aHorizonV40}
\end{figure}
\begin{figure}[htb!]
	\centering
	\includegraphics[keepaspectratio=true,scale=0.5]{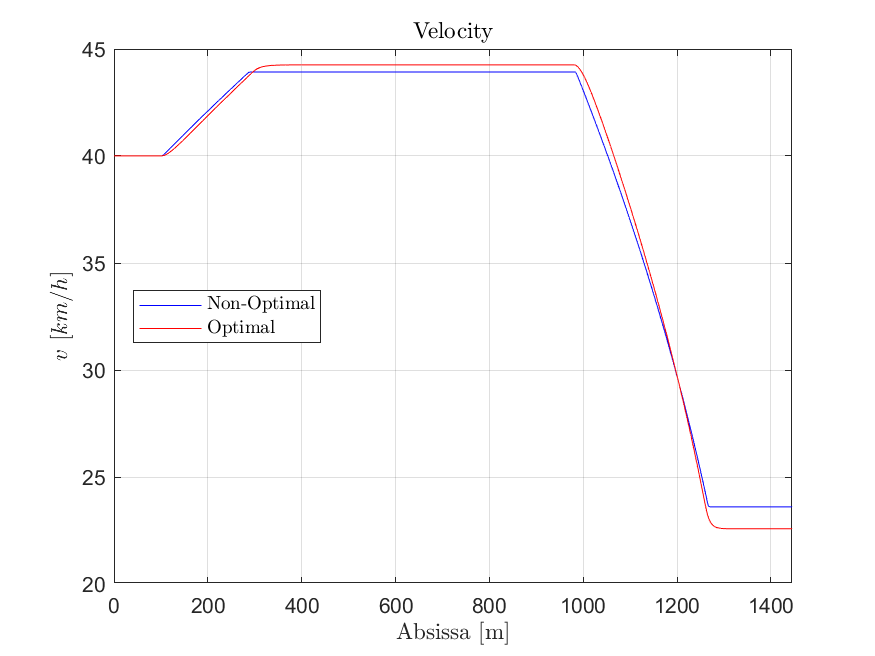}
	\caption{Optimal MTLA Test case 1 - non-optimal and optimal velocity profiles}
   \label{VelV40}
\end{figure}
\begin{figure}[htb!]
	\centering
	\includegraphics[keepaspectratio=true,scale=0.5]{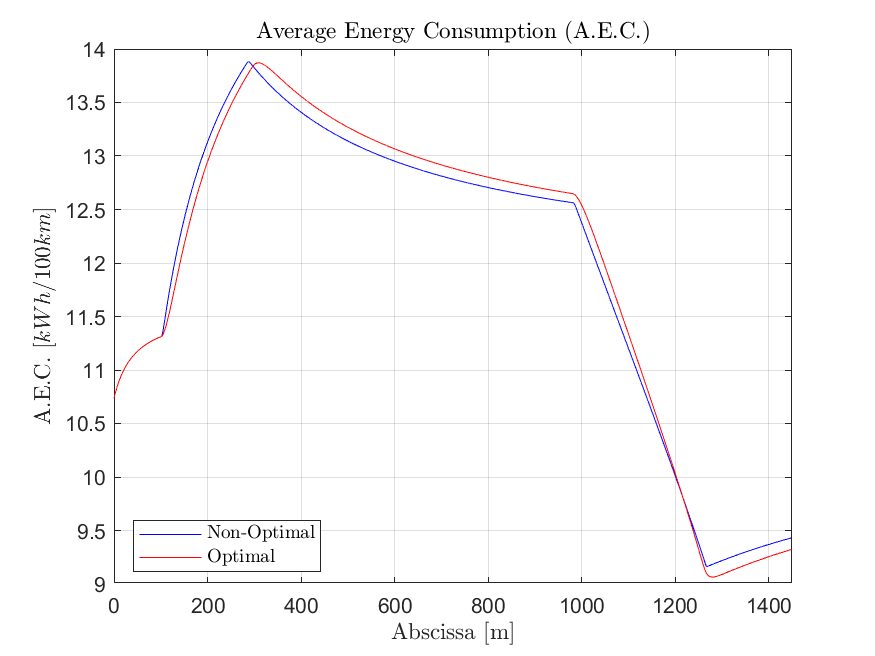}
	\caption{Optimal MTLA Test Case 1 -  non-optimal and optimal Average Energy Consumption}
    \label{OMTLWr_C1_2_AEC}
\end{figure}
Now, the non-optimal and optimal MTLA are compared for this test case. In \cref{aV40} the vehicle acceleration is plotted with respect to abscissa. Also from this plot it can be seen how the optimal strategy allows to have a more continuous acceleration profile. 
In \cref{aHorizonV40} an example of the non-optimal and optimal references along the prediction horizon is shown, the instant to which it refers is the beginning of the first acceleration phase ($100~\text{m}$ from the starting point of the simulation). It can be noticed how the predicted acceleration (optimal reference) gradually reaches the non-optimal reference acceleration. Also the velocity profiles are reported in \cref{VelV40}. Non-optimal and optimal average consumption profiles are depicted in \cref{OMTLWr_C1_2_AEC} it can be seen how, from an energy point of view, the optimal approach is not introducing significant advantages. However, a final reduction of $2 \%$ of the average energy consumption is visible at the end of the simulation. 
\section{Conclusion}
It could be seen that the following objectives were tackled:
\begin{itemize}
    \item Develop and evaluate a Traffic Light Advisor system (TLA) that relies on information shared via 5G to improve intersection viability and reduce vehicle-related emissions.
    \item Use a Model Predictive Control approach to further improve the comfort and safety of the Multiple Traffic Light Advisor system.
\end{itemize}
At first, a novel non-optimal longitudinal guidance system is developed and presented: Multiple Traffic Light Advisor (MTLA). Through a visual or acoustic warning, this system improves intersection viability and vehicle consumption by avoiding unnecessary Stop\&Go situations, and last-second braking and guides the driver to take a green wave. Unlike state-of-the-art, a uniformly accelerated motion profile in combination with constant velocity one is used in this work. This choice leads to smaller computational power requirements and makes feasible a real-time application without the necessity to improve vehicle hardware. Simulations showed that MTLA successfully fulfilled the latter objectives. Finally, an optimized version of the MTLA system is developed and tested: it optimizes the MTLA algorithm output to have a continuous, smooth and comfortable acceleration profile that triggers the warning. With respect to literature, the optimal MTLA increases comfort by minimizing jerk. Moreover, a novel algorithm to consider the passage of the vehicle only during green traffic lights is proposed.

For evaluating the system, an optimal MTLA-equipped vehicle was compared to a non-equipped one. Simulation results showed the ability of the advisor system to improve intersection viability and reduce energy consumption. Moreover, a comparison between non-optimal and optimal MTLA confirmed increased comfort due to jerk minimization.

Several future developments could be performed in order to test the proposed logic and improve them as well. Further simulations based on real scenarios could be performed to evaluate a real statistical improvement. Since in real scenarios more than one vehicle is present on the road, it would be of interest to extend the development of the TLA system by considering other vehicles. In addition, the optimal MTLA system could be adopted for autonomous driving as it would fully unveil its potential. 

\section*{Acknowledgments}
This work is done in collaboration between the Mechanical
Engineering Department, iDrive Lab of Politecnico di Milano
and Vodafone. The authors would like to thank all their
collaborating partners in the Vodafone 5G Trial in Milan:
Altran, Vodafone Automotive, Magneti Marelli, Pirelli, and
FCA. A special thanks goes to IPG for providing student
licenses that were necessary for the development and testing
of the developed work.
The Italian Ministry of Education, University and Research
is acknowledged for the support provided through the Project
``Department of Excellence LIS4.0 - Lightweight and Smart
Structures for Industry 4.0".

\section{References}

\bibliographystyle{IEEEtran}
\bibliography{mybibfile}

\end{document}